\documentclass[twoside,twocolumn,9pt]{article}
\usepackage{extsizes}
\usepackage[super,sort&compress,comma]{natbib} 
\usepackage[version=3]{mhchem}
\usepackage[left=1.5cm, right=1.5cm, top=1.785cm, bottom=2.0cm]{geometry}
\usepackage{balance}
\usepackage{mathptmx}
\usepackage{sectsty}
\usepackage{graphicx} 
\usepackage{lastpage}
\usepackage[format=plain,justification=justified,singlelinecheck=false,font={stretch=1.125,small,sf},labelfont=bf,labelsep=space]{caption}
\usepackage{float}
\usepackage{fancyhdr}
\usepackage{fnpos}
\usepackage[english]{babel}
\addto{\captionsenglish}{%
  
}
\usepackage{array}
\usepackage{droidsans}
\usepackage{charter}
\usepackage[T1]{fontenc}
\usepackage[usenames,dvipsnames]{xcolor}
\usepackage{setspace}
\usepackage[compact]{titlesec}
\usepackage{hyperref}

\usepackage{epstopdf}
\usepackage{bm}
\usepackage{amsfonts}
\definecolor{cream}{RGB}{222,217,201}
\usepackage{lineno}

\begin{document}

\pagestyle{fancy}
\thispagestyle{plain}
\fancypagestyle{plain}{
\renewcommand{\headrulewidth}{0pt}
}

\makeFNbottom
\makeatletter
\renewcommand\LARGE{\@setfontsize\LARGE{15pt}{17}}
\renewcommand\Large{\@setfontsize\Large{12pt}{14}}
\renewcommand\large{\@setfontsize\large{10pt}{12}}
\renewcommand\footnotesize{\@setfontsize\footnotesize{7pt}{10}}
\makeatother

\renewcommand{\thefootnote}{\fnsymbol{footnote}}
\renewcommand\footnoterule{\vspace*{1pt}%
\color{cream}\hrule width 3.5in height 0.4pt \color{black}\vspace*{5pt}} 
\setcounter{secnumdepth}{5}

\makeatletter 
\renewcommand\@biblabel[1]{#1}            
\renewcommand\@makefntext[1]%
{\noindent\makebox[0pt][r]{\@thefnmark\,}#1}
\makeatother 
\renewcommand{\figurename}{\small{Fig.}~}
\sectionfont{\sffamily\Large}
\subsectionfont{\normalsize}
\subsubsectionfont{\bf}
\setstretch{1.125} 
\setlength{\skip\footins}{0.8cm}
\setlength{\footnotesep}{0.25cm}
\setlength{\jot}{10pt}
\titlespacing*{\section}{0pt}{4pt}{4pt}
\titlespacing*{\subsection}{0pt}{15pt}{1pt}

\fancyfoot{}
\fancyfoot[LO,RE]{\vspace{-7.1pt}\includegraphics[height=9pt]{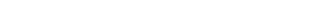}}
\fancyfoot[CO]{\vspace{-7.1pt}\hspace{13.2cm}\includegraphics{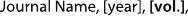}}
\fancyfoot[CE]{\vspace{-7.2pt}\hspace{-14.2cm}\includegraphics{head_foot/RF}}
\fancyfoot[RO]{\footnotesize{\sffamily{1--\pageref{LastPage} ~\textbar  \hspace{2pt}\thepage}}}
\fancyfoot[LE]{\footnotesize{\sffamily{\thepage~\textbar\hspace{3.45cm} 1--\pageref{LastPage}}}}
\fancyhead{}
\renewcommand{\headrulewidth}{0pt} 
\renewcommand{\footrulewidth}{0pt}
\setlength{\arrayrulewidth}{1pt}
\setlength{\columnsep}{6.5mm}
\setlength\bibsep{1pt}

\makeatletter 
\newlength{\figrulesep} 
\setlength{\figrulesep}{0.5\textfloatsep} 

\newcommand{\topfigrule}{\vspace*{-1pt}%
\noindent{\color{cream}\rule[-\figrulesep]{\columnwidth}{1.5pt}} }

\newcommand{\botfigrule}{\vspace*{-2pt}%
\noindent{\color{cream}\rule[\figrulesep]{\columnwidth}{1.5pt}} }

\newcommand{\dblfigrule}{\vspace*{-1pt}%
\noindent{\color{cream}\rule[-\figrulesep]{\textwidth}{1.5pt}} }

\makeatother

\twocolumn[
  \begin{@twocolumnfalse}
{\includegraphics[height=30pt]{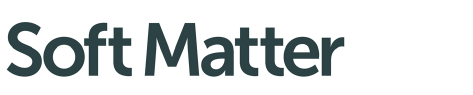}\hfill\raisebox{0pt}[0pt][0pt]{\includegraphics[height=55pt]{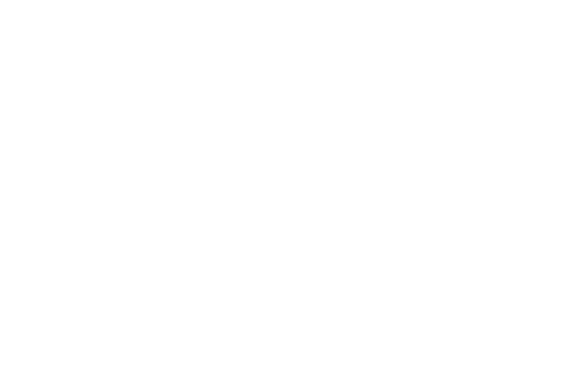}}\\[1ex]
\includegraphics[width=18.5cm]{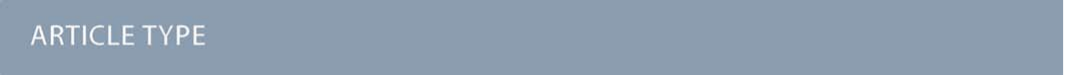}}\par
\vspace{1em}
\sffamily
\begin{tabular}{m{4.5cm} p{13.5cm} }

\includegraphics{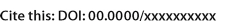} & \noindent\LARGE{\textbf{Confinement twists achiral liquid crystals and causes chiral liquid crystals to twist in the opposite handedness: Cases in and around sessile droplets$^\dag$}} \\
\vspace{0.3cm} & \vspace{0.3cm} \\

 & \noindent\large{Jungmyung Kim\textit{$^{a}$} and Joonwoo Jeong$^{\ast}$\textit{$^{a}$}} \\

\includegraphics{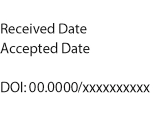} & \noindent\normalsize{We study the chiral symmetry breaking and metastability of confined nematic lyotropic chromonic liquid crystal (LCLC) with and without chiral dopants. The isotropic-nematic coexistence phase of the LCLC renders two confining geometries: sessile isotropic(I) droplets surrounded by the nematic(N) phase and sessile nematic droplets immersed in the isotropic background. 
In the achiral system with no dopants, LCLC's elastic anisotropy and topological defects induce a spontaneous twist deformation to lower the energetic penalty of splay deformation, resulting in spiral optical textures under crossed polarizers both in the I-in-N and N-in-I systems. 
While the achiral system exhibits both handednesses with an equal probability, a small amount of the chiral dopant breaks the balance. 
Notably, in contrast to the homochiral configuration of a chirally doped LCLC in bulk, the spiral texture of the disfavored handedness appears with a finite probability both in the I-in-N and N-in-I systems.
We propose director field models explaining how chiral symmetry breaking arises by the energetics and the opposite-twist configurations exist as meta-stable structures in the energy landscape. These findings help us create and control chiral structures using confined LCs with large elastic anisotropy.
} \\

\end{tabular}

 \end{@twocolumnfalse} \vspace{0.6cm}
]

\renewcommand*\rmdefault{bch}\normalfont\upshape
\rmfamily
\section*{}
\vspace{-1cm}


\footnotetext{\textit{$^{a}$Department of Physics, Ulsan National Institute of Science and Technology, Ulsan, Republic of Korea. E-mail: jjeong@unist.ac.kr}}

\footnotetext{\dag~Electronic Supplementary Information (ESI) available: [details of any supplementary information available should be included here]. See DOI: 10.1039/cXsm00000x/}



Chiral structures are ubiquitous, ranging from molecules to astrophysical phenomena. Understanding the mechanism underlying this mirror symmetry breaking is an active field of investigation. Simultaneously, the control and separation of chiral structures are crucial in diverse applications \cite{Green1989, Brunsveld2002, Pakhomov2003, Hough2009, Kang2013}. For instance, chiral molecules can be toxic in unwanted enantiomeric forms \cite{Smith2009}, and these chiral objects as building blocks or dopants can determine their assembled structures' chirality. Additionally, external fields, including shear and confinement, are often employed to control chiral structures \cite{Petit-Garrido2009, Tortora2011, Pairam2013, Jeong2014a}.

Liquid crystal (LC) is the quintessential model system for studying chiral structures \cite{Kitzerow2001, Dierking2014, Salamo_czyk_2019, Abberley_2018, Lewandowski_2020, Wang2020, Majewska_2022}. In particular, chiral nematic LCs, also known as the cholesteric phase, have been studied extensively. They are omnipresent in nature and deployed in various applications \cite{Huang2006, Sharma2009, Han2010, Coles2010, Beeckman2011, Mitov2017, Hare2020}. In the cholesteric phase, LC mesogens twist along the helical axis, defining its handedness --- right- or left-handed --- and helical pitch, which are often determined by the intrinsic chirality of the mesogen or dopants. Thus, one may design chiral structures by choosing mesogens or the dopant's type and concentration at a given temperature.

The boundary conditions and elasticity of confined LCs also play a vital role in determining their chiral structures \cite{Press1974, Kitzerow1996, Ambrozic1996, Ambrozic1999, Tortora2011, Pairam2013, Nych2014, Jeong2014a, Davidson2015, Nayani2015, Jeong2015, Peng2015, Guo2016, Ryu2016, Zhou2016, Serra2016, Javadi2018, Ellis2018, Lee2019, Palacio-Betancur2023}.
The confining geometry and LC surface anchoring at the interface between LC and the surrounding phase can result in unexpected structures that differ from the ones in bulk.
For example, confinement may lead to an apparent helical pitch different from the bulk pitch\cite{Smalyukh2002, Eun2021}.
Furthermore, achiral LCs under confinement, \textit{e.g.}, droplets and cylinders, can form chiral structures \cite{Volovik1983, Williams1986, DRZAIC1999, Tortora2011, Pairam2013, Jeong2014a, Nych2014, Davidson2015, Nayani2015, Jeong2015, Ignes-Mullol2016, Dietrich2017} and vice versa \cite{Guo2016}.

Confinements may afford multiple meta-stable structures \cite{Sec2012, Yoshioka2016, Eun2021, Li2023}, often resulting in chiral structures having the opposite handedness to their bulk ones, of which the handednesses and helical pitch are determined by the chirality of the mesogens and dopants. For example, the twist handedness can be reversed locally \cite{Ackerman2016, Pollard2019, Li2023}, or domains of different handedness coexist \cite{Raynes2006, Eun2019, Dietrich2021}, resulting in heterochiral structures despite their energetic penalty. These confinement effects involving surface anchoring, elastic anisotropy, saddle-splay elasticity, topological defects, and metastability, have advanced our understanding of the partially ordered matter and its applications.

Here, we report cases in and around sessile droplets where chiral nematic LC can exhibit a meta-stable structure of the disfavored handedness, \textit{i.e.}, opposite to the handedness in bulk, because of boundary conditions.
We use achiral and chiral nematic lyotropic chromonic LCs to study their director configuration in and around sessile droplets on an anchoring-controlled substrate.
Our experiments and numerical calculations demonstrate the roles of confinement and elastic anisotropy in the chiral symmetry breaking of achiral LC.
Moreover, we elucidate how chiral LC can afford chiral structures of the opposite handedness as meta-stable states, thanks to confinements.

\section*{Materials and Methods}
\subsection*{Materials}
Sunset Yellow FCF (SSY, a purity of 90\%) was purchased from Sigma-Aldrich,  and we purified using the conventional method \cite{Horowitz2005} because the impurities matter in their phase behavior \cite{Eun2020} and, presumably, elastic properties.
Brucine sulfate heptahydrate (BSH) as a chiral dopant for SSY was purchased from Sigma-Aldrich and used without purification.
Mixing deionized water, SSY, and BSH, we prepared an aqueous solution of 30.0\% (wt/wt) SSY with 0 to 0.4\% (wt/wt) of BSH as achiral (without BSH) or chiral nematic LC solution at room temperature.

\subsection*{Sample Preparation}
We coated glass slides and coverslips with parylene \cite{Jeong2014} and used them as substrates for the homeotropically aligned cell.
The substrates were placed in a parylene coating machine (LAVIDA 110; Femto Science) with 500 mg of parylene-N (diX N, Daisan Kasei), and the temperatures for vaporizer, pyrolyzer, and deposition of parylene were $150^{\circ}\mathrm{C}$, $690^{\circ}\mathrm{C}$, and room temperature, respectively.
The estimated thickness of the parylene film is approximately 400 nm \cite{Lee2011}.
Utilizing Kapton films of 25 $\mu$m thickness on the surface-treated slide as spacers, we dropped 10 $\mu$l of LC solution on the parylene-coated slide and quickly covered the specimen with the coated coverslip.
The edges of the sandwich cell were sealed with epoxy glue to minimize the drying of the aqueous LC solution.
The sandwiched nematic SSY initially shows the transient stripe texture, but after several tens of minutes, the stripe texture is replaced by a full homeotropically aligned domain \cite{Jeong2014}.
However, when the dopant concentration surpasses 0.4\% (wt/wt), the sandwiched nematic SSY no longer exhibits the homeotropic alignment but the fingerprint texture, not providing the proper initial condition for our droplet experiments, as shown in Fig. S1.

\subsection*{Optical Observation}
We observed the sandwich cell with a polarized optical microscope (Olympus BX53-P) equipped with linear polarizers, full waveplate (optical path difference = 550 nm; Olympus), a temperature-controlled stage (T95-PE120; Linkam Scientific Instruments), a CCD camera (INFINITY3-6URC; Lumenera), and a LED lamp (LED4D067; Thorlabs) providing quasi-monochromatic illumination at 660 nm wavelength (FWHM = 25 nm).
We increased the temperature, $1^{\circ}\mathrm{C}/$min, of the aligned LC cell to make the LC enter the isotropic-nematic coexistence phase at $45^{\circ}\mathrm{C}$.
Repeating the heating-cooling sequence, we collected multiple polarized optical microscopy (POM) images from the same location of the same sample. Note that we employ the slow heating rate to control the droplet locations, \textit{i.e.}, only at the bottom, and propose experiments varying the rate as future work.

\subsection*{Numerical Calculations}
The total elastic free energy $F$ is the volume integral of the energy density $f$ given by 
\begin{gather}
f=\frac{1}{2} [K_{1}(\nabla \cdot \mathbf n)^2+K_{2}(\mathbf n \cdot \nabla \times \mathbf n + 2\pi/p_0)^2 \nonumber\\
+K_{3}(\mathbf n \times \nabla \times \mathbf n)^2],
\label{eq:freeE}
\end{gather}
where $K_{1}$, $K_{2}$, and $K_{3}$ are elastic moduli of splay, twist, and bend deformations, respectively. Note that we neglect a saddle-splay elasticity term with $K_{24}$ because it does not affect the energetics when the homeotropic anchoring is strong (at the substrates), and the degenerate planar anchoring is imposed on the surface with the same principal curvatures (the surface of the hemispherical droplet). $p_0$ is the helical pitch of the chiral nematic phase and is inversely proportional to the concentration of a chiral dopant.
Assuming that the elastic moduli remain similar in the nematic phase of the coexistence and the moduli do not depend on the dopant concentration, we use $K_{1}:K_{2}:K_{3}=1:0.1:1$ for the 30.0\% (wt/wt) SSY \cite{Zhou2012} and $p_{0} = 1/(C*HTP)$ where $C$ is BSH concentration in \% (wt/wt) and $HTP$ is helical twisting power, 0.123 $\pm$ 0.007 $\mu$m$^{-1}$ (see Fig. S2 for details). Note that we employ a constant $p_0$ with no spatial dependency, assuming a uniform distribution of the chiral dopant. We believe this assumption deserves further investigation, considering the fine structure of the topological defect cores of LCLCs \cite{Zhou2017}. 
For a given director field $\mathbf{n}$, we perform a numerical integration $F=\int f dv$ over domain $\mathbb{D}$ to draw the energy landscape such as in Fig. \ref{fig:energymapIN}(a).
For the I-in-N case, \textit{i.e.}, the sessile isotropic droplet surrounded by the nematic phase, we adopted $\mathbb{D}=\{(x,y,z)\mid x^2+y^2\leq (5R)^2, 0\leq z\leq 5R, x^2+y^2+z^2\geq R^2, x^2+y^2+(z-R)^2\geq R_d^2\}$ where $R$ is the radius of the hemispherical droplet.
To facilitate the convergence of the numerical integration, we set a cutoff radius $R_d=10^{-3}R$ around the point defect at the top of the droplet; $R_d=10^{-3}R$ corresponds to 10 nm, which is of the same order of the persistence length of the SSY aggregates \cite{Zhou2012} in a droplet of 10 $\mu$m radius.
For the N-in-I case, we used $\mathbb{D}=\{(x,y,z)\mid 0\leq z, x^2+y^2+z^2\leq (R)^2, x^2+y^2+(z-R)^2\geq R_d^2\}$ and $R_d=10^{-3}R$.
After identifying the director configurations corresponding to local minima in the energy landscape of each director fields model, we simulated their POM textures using Jones calculus\cite{Jeong2014a} and compared the results with experimental observations.

\section*{Results and Discussion}
\subsection*{Sessile Isotropic Droplets Surrounded by Nematic Phase}

\begin{figure}[t]
\centering
\includegraphics{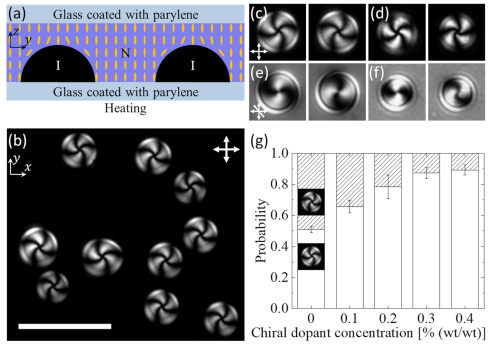}
\caption{
\label{fig:pom}
Spiral textures in the isotropic-nematic coexistence phase of Sunset Yellow under confinement.
\textbf{(a)} Schematic diagram of the sample from the side view. The isotropic-nematic coexistence phase is sandwiched between two parylene-coated substrates.
Sessile droplets of the isotropic(I) phase surrounded by the nematic(N) sit on the heated bottom substrate.
Yellow rods in the nematic phase sketch the directors.
\textbf{(b)} Spiral textures observed with crossed polarizers from the top view. White double arrows represent the pass axes of the polarizer and analyzer. The scale bar is 50 $\mu$m.
\textbf{(c)-(d)} Representative spiral textures having different handedness (c) without chiral dopant and (d) with 0.4\% (wt/wt) chiral dopant.
The width and height of the images are 30 $\mu$m, respectively.
\textbf{(e)-(f)} Spiral textures of (c) and (d) with a full waveplate.
(g) Appearance probability of the spiral textures having different handedness according to the chiral dopant concentration.
As the concentration increases, a certain handedness becomes dominant, but the other handedness exists even at the highest dopant concentration investigated.}
\end{figure}

We observe chiral symmetry breaking of achiral nematic LC around sessile isotropic liquid droplets.
Figure \ref{fig:pom} shows our lyotropic chromonic LC, SSY, at its isotropic-nematic coexistence phase confined between two parallel substrates imposing homeotropic anchoring.
When a homeotropically aligned nematic SSY enters the coexistence phase upon a temperature increase, as shown in Fig. \ref{fig:pom}(a), sessile droplets of the isotropic phase appear at the bottom substrate.
Because the temperature-controlling Peltier plate contacts the bottom substrate, the isotropic droplets nucleate from the bottom substrate.
We make and observe the droplets only at the bottom substrate and find the contact angle at the three-phase boundary is close to 90 degrees, as shown in ESI\dag.
We also confirm no memory effect by comparing the location, size, and handedness of droplets in different images.
The sessile isotropic droplets deform the homeotropically aligned nematic director field, and the deformed regions exhibit spiral textures under POM, as shown in Fig. \ref{fig:pom}(b)-(f). The appearance of the chiral spiral texture, despite the absence of chirality in SSY, is notable.
Non-deformed nematic regions remain dark under crossed polarizers because of the homeotropic alignment.

We classify the spiral textures' handedness, investigate their appearance probability, and find that each handedness is equally probable.
In the right-handed spiral texture (RS), the dark line turns right while going from the center to the boundary.
In three independent experiments, each collecting approximately 270 droplets, the average probability of observing the RS texture is 0.509 $\pm$ 0.021, where the error indicates the standard deviation of the three measurements.
This half-and-half probability is consistent with the achiral nature of SSY, \textit{i.e.}, no preference for a specific handedness.

With a chiral dopant added, the spiral texture prefers a specific handedness.
As we increase the concentration of a dopant, BSH, the appearance probability of RS increases and excels left-handed spiral texture (LS)'s, as shown in Fig. \ref{fig:pom}(g).
For example, with 0.4\% (wt/wt) BSH added to 30\% (wt/wt) SSY, almost 90\% of the spiral texture becomes RS.
It is unexpected that 10\% remains as the energetically unfavorable LS at our maximum dopant concentration because a chirally doped LC in bulk would exhibit a homochiral director configuration with a well-defined helical pitch \cite{Lee1982, McGinn2013, Peng2015, Ogolla2017, Shirai2018, Ogolla2019}. On the other hand, we propose as future work the investigation of the appearance probability at very low concentration, \textit{i.e.}, between 0 and 0.1\% (wt/wt), to find the critical concentration for the chiral symmetry breaking \cite{Eun2019}.

To elucidate the observed chiral symmetry breaking in achiral SSY and metastability in chiral SSY, we propose a director field model of nematic LC around the isotropic droplet.
We employ this ansatz approach instead of the numerical simulation adopted in a similar geometry \cite{Nych2014} for the following reason.
The numerical simulation using the Q-tensor can find the energy-minimizing director configurations that better match experimental observations, although parameter choices for LCLCs require further studies. 
The director configurations found by the model with a finite number of parameters are undoubtedly approximated ones.
However, the model approach often can be useful, enabling us to map the energy landscape of the problem systematically.
Then, we can not only identify local minima leading to meta-stable configurations but also rationalize why and how the minima exist, which will be the main focus of this paper. 

\begin{figure}[t]
\centering
\includegraphics{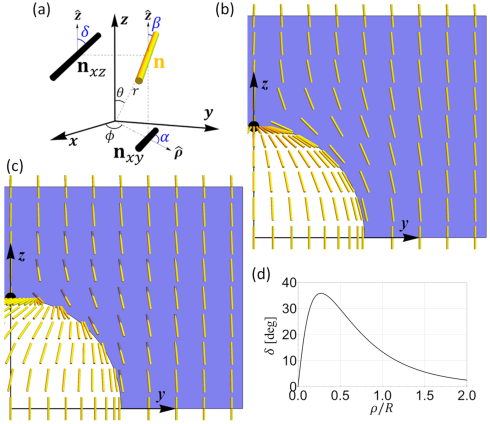}
\caption{\label{fig:director1}
Director configurations of a sessile isotropic droplet surrounded by nematic phase.
\textbf{(a)} A coordinate system to represent the director configuration.
A yellow rod is the director at the position $\mathbf{r}$, and two black rods are the projection of the yellow director on $x-y$ plane and $x-z$ plane, respectively.
$\alpha$, $\beta$, and $\delta$ are the angles between the coordinate axes and the director or the projected directors.
\textbf{(b)} Director visualization of the planar quadrupole configuration. The directors are shown as yellow rods on the surface of the droplet and on the $y-z$ plane. The black dot at the top of the droplet is the point defect.
\textbf{(c)} One example of the twisted planar quadrupole configuration.
\textbf{(d)} Twist angle $\delta$ profile of (c) as a function of the normalized radius $\rho/R$ in the cylindrical coordinate along the $\rho$ axis at $z/R=1.1$.
}
\end{figure}

The first assumption of our model is that the isotropic sessile droplet in the model has a hemispherical shape, as shown in Fig. \ref{fig:pom}(a). This assumption is supported by the near 90-degree contact angle we estimate at the three-phase boundary.
It is surrounded by the nematic phase satisfying the boundary conditions: an infinitely strong homeotropic anchoring both on the top and bottom substrates and a degenerate planar anchoring on the hemispherical droplet's surface.
The hemispherical geometry and anchoring conditions differentiate our work from the chiral bipolar configuration reported in Nych \textit{et al}. and Uzunova and Pergamenshchik \cite{Uzunova2011, Nych2014}. Additionally, the previous works study only the nematic phase, but our system investigates both achiral and chiral nematic phases.
We parameterize the director field $\mathbf{n}$ with two spatially varying angles, $\alpha$ and $\beta$.
\begin{gather}
(n_x, n_y, n_z) = (\sin{\beta}\cos{(\phi+\alpha)},\sin{\beta}\sin{(\phi+\alpha)},\cos{\beta}).
\label{eq:director}
\end{gather}
The equation describes the Cartesian components $(n_x, n_y, n_z)$ of a director $\mathbf{n}$ at a position ($r$, $\theta$, $\phi$) in the spherical coordinate.
For the angle $\beta$ between a director and $z$-axis shown in Figure \ref{fig:director1}(a), we adopt the bipolar planar quadrupole configuration \cite{Uzunova2011}.
\begin{gather}
\beta=2\theta-\tan^{-1}\left(\frac{(r/R)^{3/2}\sin{\theta}}{(r/R)^{3/2}\cos{\theta}-1}\right)\nonumber\\-\tan^{-1}\left( \frac{(r/R)^{3/2}\sin{\theta}}{(r/R)^{3/2}\cos{\theta}+1}\right).
\label{eq:beta}
\end{gather}
This ansatz adopted Eq. 6 from Uzunova and Pergamenshchik\cite{Uzunova2011} satisfies our boundary conditions, \textit{i.e.}, the homeotropic alignment at the far field and the degenerate planar alignment on the hemispherical droplet's surface, resulting in a point defect at the droplet's pole, also known as a boojum. Note that our sessile-droplet geometry has one pole at the top of the droplet.

\begin{figure}[t]
\centering
\includegraphics{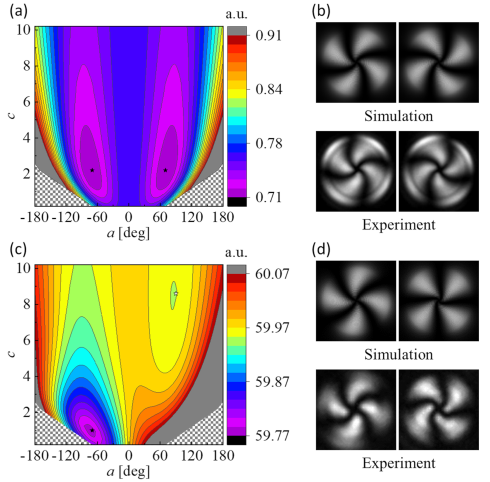}
\caption{\label{fig:energymapIN} Elastic free energy landscape of the I-in-N system according to the model parameter $a$ and $c$ and their comparison with experimental data.
\textbf{(a)} The color-coded energy landscape of the achiral nematic SSY with no chiral dopant.
Two filled star symbols indicate two degenerate global minima. The patterned regions at the bottom left and right are excluded because they violate the boundary conditions considerably.
\textbf{(b)} Comparison of experimentally observed spiral textures with Jones-calculus-simulated POM textures of the director configurations located at two global minima of the energy landscape shown in (a).
\textbf{(c)} The color-coded energy landscape of the chiral nematic SSY with 0.4\% (wt/wt) BSH.
The global minimum is indicated by a filled symbol, and the local minimum at the top-right region is shown with an empty star.
\textbf{(d)} Comparison of experimental images with simulated POM textures based on the minima in (c).
Left and right column of POM images represent global and local minimum, respectively.
}
\end{figure}

We hypothesize that the experimentally observed spiral POM textures result from a twisted director configuration. We add twist deformation to the planar quadrupole configuration by introducing the angle $\alpha$ in Eq. \ref{eq:director}.
As shown in Fig. \ref{fig:director1}(a), $\alpha$ is the angle between the radial direction $\hat{\rho}$ in the cylindrical coordinate and the director's projection onto the $x-y$ plane.
If $\alpha=0$, the director field becomes the planar quadrupole configuration with no $\hat{\phi}$-component.

Our ansatz adopts $\alpha$ as a power-law decaying function of $\rho$, satisfying the cylindrical symmetry and the far-field boundary condition. 
We choose this decaying profile because the splay cancellation by the twist deformation near a topological defect is a well-known mechanism for the chiral symmetry breaking in LCLCs \cite{Jeong2014a}; see ESI for the detailed rationalization of the ansatz.
Namely, $\alpha$ decreases along the radial direction from the $z$ axis, \textit{i.e.}, from the boojum, and satisfies $\alpha(\rho \to \infty)\to 0$, \textit{i.e.}, the deformation-free homeotropic alignment at the infinity.
\begin{gather}
\alpha=a \left( \frac{1}{1+(\rho/R)} \right) ^{c},
\label{eq:acpara}
\end{gather}
Two constants $a$ and $c$ determine the decaying $\alpha(\rho)$'s profile and how the twist angle $\delta$ defined in Fig. \ref{fig:director1}(a) varies spatially.
Fig. \ref{fig:director1}(d) is a representative $\delta$-$\rho$ profile for given $a$ and $c$. A large $a$ results in the large peak value of $\delta$, and a large $c$ leads to a fast decay as $\rho$ increases.

The degenerate planar anchoring of the planar quadrupole configuration at the droplet surface can be violated when we adopt $\alpha \neq 0$ for the twist deformation.
Varying the parameters $a$ and $c$, we investigate the maximum polar deviation angle of the surface directors from the tangential plane at the droplet surface.
Then, we exclude the parameter regimes having the maximum deviation angle greater than 20 deg from the region of interest, only allowing a slight violation of the planar anchoring condition at the isotropic-nematic interface.

By numerically calculating the total elastic free energy $F$ of the above-mentioned director field model according to $a$ and $c$, we find the director configurations at local energy minima, which agree well with experimental observations.
See the Methods section for the details of the elastic free energy calculation.
Figure \ref{fig:energymapIN}(a) shows the energy landscape of the achiral system, highlighting the local minima.
In this achiral system, the energy landscape is symmetric about the axis of $a=0$, \textit{i.e.}, $F(a) = F(-a)$; the sign change of the angle $\alpha$ means the flipping of the spiral handedness.
We locate $(a, c)$ giving the two degenerate global minima of the energy landscape, simulate the corresponding POM textures using Jones calculus, and compare them with the experimental results, as shown in Fig. \ref{fig:energymapIN}(b).
The simulated textures well reproduce the spiral textures of different handednesses.
Additionally, two degenerate global minima are consistent with the observation that the spiral textures of different handedness appear with equal probabilities in the achiral system.

The elastic anisotropy of the nematic SSY is responsible for two global minima at non-zero $a$ and $c$, resulting in the two degenerate twisted ground states.
Fig. S3 shows the energy difference between our twisted ground-state configuration and the twist-free planar quadrupole configuration.
It indicates that the relatively cheap twist deformation cancels out the splay deformation near the topological defect, resulting in a net decrease in the total elastic free energy.
This scenario is similar to the cases in the literature, where the achiral SSY under confinement breaks its chiral symmetry because of the very small $K_2$ compared to other elastic moduli \cite{Jeong2014a, Nayani2015, Jeong2015, Eun2019}.
Comparison of energy landscapes at different $K_{2}$s corroborates the role of elastic anisotropy in the chiral symmetry breaking.
Our close investigation of the energy landscape reveals two degenerate global minima exist when $k_{2} = K_{2}/K_{3} \leq 0.68$; See ESI\dag for the determination of the critical $k_2$ for the chiral symmetry breaking.

With chiral dopants added, as shown in Fig. \ref{fig:energymapIN}(c), the energy landscape becomes asymmetric but still retains two minima. 
This is consistent with our experimental observation that the spiral textures of both handednesses appear.
Because $p_0$ in Eq. \ref{eq:freeE} is no longer zero, the director configuration energetically favors a twist deformation of a given helical pitch and handedness imposed by the chiral dopant, making the energy landscape asymmetric.
Notably, two global minima in the achiral system shown in Fig. \ref{fig:energymapIN}(a) become one global minimum and one local minimum in the chiral system shown in Fig. \ref{fig:energymapIN}(c).
The director configuration of the preferred handedness has the lower energy and corresponds to the global minimum, which is a more frequently observed spiral texture in our experiments.
For instance, 0.4\% (wt/wt) of chiral dopant gives right-handed spiral texture handedness with the probability 0.894 $\pm$ 0.032 out of 794 droplets.
Intriguingly, the director configuration of the disfavored handedness exists with a finite probability, even though we add the chiral dopant to make the configuration favor a particular handedness. In fact, in the calculated energy landscape, there exists the local minimum with higher energy than the global minimum; See the top-right region of Fig. \ref{fig:energymapIN}(c).

We find that a constraint from the symmetry and the far-field boundary condition results in the appearance of both handednesses.
As shown in Fig. \ref{fig:director1}(c), the directors along the rotational axis, \textit{i.e.}, $z-$axis, are aligned vertically because of the symmetry.
Starting from this symmetric axis, the directors twist along the radial direction $\hat{\rho}$ to lower the total elastic free energy in both achiral and chiral cases.
However, to satisfy the far-field boundary condition, \textit{i.e.}, the uniform homeotropic alignment, the director should untwist to be vertical again unless the centrosymmetric directors twist by 180 degrees.
In other words, as shown in Fig. \ref{fig:director1}(d), the system should have both twist-handedness, \textit{i.e.}, the positive and negative slopes in $\delta$, regardless of the favored handedness of the system.
This constraint justifies that, even in highly doped specimens preferring one-handedness and paying high energy penalty for the other-handedness, the disfavored spiral texture can be observed experimentally as a local minimum in the energy landscape.
With different symmetry and boundary conditions allowing the homochiral twist deformation, one may expect chiral LCLCs to exhibit the homochiral spiral texture. We test this idea in the following section with the inverted system.

\subsection*{Sessile Nematic Droplets Surrounded by Isotropic Phase}

\begin{figure}[t]
\centering
\includegraphics{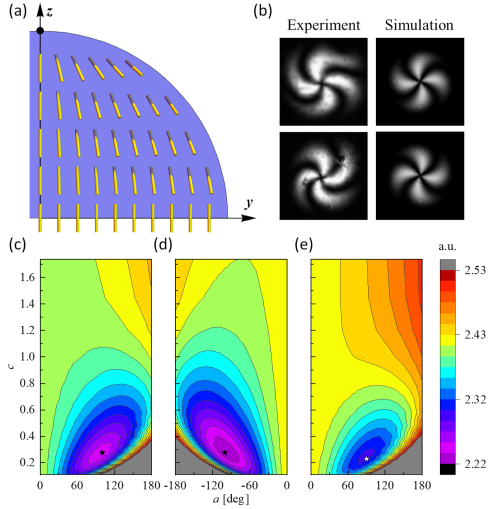}
\caption{\label{fig:ni} Schematic, spiral textures, and energy landscape of the N-in-I system.
\textbf{(a)} Director visualization of a representative twisted configuration in the sessile nematic droplet.
The directors are shown as yellow rods on the $y-z$ plane.
The black dot at the top of the droplet is the point defect.
\textbf{(b)} The comparison of experimental POM textures with simulated textures based on the energy-minimizing configurations of the energy landscape.
\textbf{(c)} The color-coded energy landscape of the achiral nematic SSY according to the model parameters $a$ and $c$ at $b=2.51$. The symmetry landscape with the negative $a$ is not shown. The filled symbol indicates one of two degenerate global minima.
\textbf{(d-e)} The color-coded energy landscape of the chiral liquid crystal with 0.1\% (wt/wt) BSH.
(d) shows the global minimum as the filled symbol in the $a-c$ plane at $b=2.09$. (e) shows the local minimum at $b=2.51$.
}
\end{figure}

We observe the spiral textures again from sessile nematic LCLC droplets surrounded by the isotropic phase, \textit{i.e.}, N-in-I, of which the boundary conditions allow homochiral configurations.
Figure \ref{fig:ni}(b) shows the POM textures.
In contrast to the I-in-N case at the temperature just above the nematic-to-coexistence phase transition temperature, the temperature of the N-in-I sample is at the one just below the coexistence-to-isotropic phase transition temperature.
The isotropic regions grow and merge during the heating, and nematic regions eventually become sessile droplets surrounded by the isotropic phase.
We find that sessile droplets surrounded by air can implement the same boundary conditions: the near hemispherical shape and degenerate planar anchoring at the air-nematic interface.
The spiral textures result from the twist deformation of the nematic phase confined within the sessile droplets. 
The absence of the far-field boundary condition, such as the homeotropic alignment in the I-in-N case, allows homochiral configurations.
This chiral symmetry breaking and the energetics are similar to those in the suspended spherical droplets \cite{Jeong2014a}.
The LCLC's cheapest twist deformation cancels out the high elastic energy penalty around the surface point defect, \textit{i.e.}, the black dot in Fig. \ref{fig:ni}(a).
As expected, in the achiral system with no chiral dopant, we observe both handednesses in the spiral texture with equal probabilities: the ratio of RS to LS is 9 to 8.

To our surprise, the sessile nematic droplets of chirally doped LCLC do not exhibit homochirality, even though the geometry and boundary conditions allow homochiral configurations.
When 0.1\% (wt/wt) of BSH is added, 41 out of 43 are left-handed.
Namely, we observe the disfavored handedness experimentally, even though they are rarely observed.
When we increase the dopant concentration to 0.2\% (wt/wt), all droplets exhibit the homochiral configuration. 
Please note that we observe the same metastability in the N-in-air system; the data is not shown here because the sessile-droplet shapes surrounded by the air are not regular enough to be a well-defined confinement.

To investigate further the energetics of the meta-stable configuration having the disfavored handedness, we construct a parameterized director field model for the hemispherical sessile droplet, which is inspired by the twisted bipolar director configuration of the spherical droplet \cite{Jeong2014a}. As in the I-in-N case, we assume the hemispherical shape because of the near 90-deg contact angle, as shown in ESI\dag.
The director $\mathbf{n}$ with a parameter $\gamma$ is 
\begin{gather}
\mathbf{n}=\mathbf{n}_{\mathrm{bipolar}}\cos{\gamma}+\mathbf{n}_{\mathrm{concentric}}\sin{\gamma},
\end{gather}
where $\mathbf{n}_{\mathrm{bipolar}}$ and $\mathbf{n}_{\mathrm{concentric}}$ are the director field model of the bipolar and concentric configurations used in Jeong \textit{et al}\cite{Jeong2014a}.
We introduce the parameter $\gamma$ to describe the azimuthal component of the director experiencing a twist deformation.
If $\gamma = 0$, the director field becomes the bipolar configuration, $\mathbf{n}=\mathbf{n}_{\mathrm{bipolar}}$, with no twist deformation. Note that we use only the hemispherical part of the bipolar configuration, resulting in one pole.
With $\gamma\neq 0$, directors twist as shown in Fig. \ref{fig:ni}(a).
Assuming $\gamma = a (z/R)^{b} (\rho/R)^{c}$ as a power-law function of $z$ and $\rho$ with parameters $a$, $b$, and $c$, we numerically calculate total elastic free energy $F$ according to ($a$, $b$, $c$). It is noteworthy that $z^{b}$ makes the model satisfy the homeotropic anchoring condition at the bottom substrate.
Moreover, the splay cancellation near the defect inspires the decaying along $\rho$. The farther the director is away from the defect, the less twist deformation is required to lower the energy.

We find that both achiral and chiral models can have two minima in the energy landscape of the parameter space $(a, b, c)$.
It is no surprise that the energy landscape of the achiral degenerate system with $p_0 = 0$ renders the two energy minima with the same energy, and these two configurations correspond to each spiral texture of different handedness shown in the experiment.
As shown in Fig. \ref{fig:ni}(b), the Jones calculus-simulated POM textures approximately match our experimental observation.
However, the existence of the local minimum configuration having the disfavored handedness, which is experimentally observed, is unexpected. It is not straightforward to imagine why the directors twist in an energetically unfavorable handedness and remain stable enough to be observed experimentally, although the boundary conditions afford homochiral configurations.

Our investigation of the energetics reveals that the elastic anisotropy is responsible for the metastability in the chiral system, as in the chiral symmetry breaking of the achiral system.
First, the energy landscape assuming the one-constant approximation, \textit{i.e.}, $K_1 = K_2 = K_3$, shows only one minimum exhibiting the twisted director configuration of the favorable handedness.
However, with $K_1 = K_3 = 10~K_2$, in addition to the global minimum, there exists a local minimum having the opposite handedness.
To rationalize the existence of the local minimum, we investigate the energy landscape in detail.
As shown in Fig. S4, in the vicinity of the local minimum, the increase in $b$ of $\gamma = a (z/R)^{b} (\rho/R)^{c}$ results in the increase of total splay energy within the droplet.
But, both total twist and bend energy decrease, leading to a local minimum at a finite $b$.
This trade-off results from the fact that the director configurations should satisfy the boundary conditions imposed by the confinement.
Namely, because of the constraints by the boundary conditions, each deformation cannot change independently, but changes in a certain deformation affect the other deformation modes.
It is the elastic anisotropy that leads to a local minimum in the trade-off situation.
Notably, with $K_1 = K_2 = K_3$ instead of $K_1 = K_3 = 10~K_2$, \textit{i.e.}, the decrease of total twist and bend energy upon the increase in $b$ dominates over the increase of total splay, giving no local minimum but a monotonic decrease.
In summary, a confinement-induced energy trade-off with the small $K_2$, or large enough elastic anisotropy in general, makes this local minimum reside even in the energy-landscape region of disfavored handedness.

\section*{Conclusions}
We study the roles of confinement and elastic anisotropy in the chiral symmetry breaking of achiral LCLCs and the metastability of chiral LCLCs.
The broad isotropic-nematic coexistence phase of LCLC enables us to study sessile isotropic droplets immersed in an aligned nematic phase and vice versa.
Our director field models reproduce the spiral textures observed by POM and elucidate how their handedness population depends on the chiral dopant concentrations.
Particularly, energy landscape calculations reveal that the spirals of the disfavored handedness can exist as a local minimum.
Regardless of the boundary conditions, \textit{i.e.}, both in I-in-N and N-in-I cases, the large elastic anisotropy $K_2\ll  K_1 = K_3$ is responsible for local minimum.

Looking forward, we should be able to utilize our findings and models for further studies of confined chiral chromonic LCs and their applications. 
We learn that energy-landscape study can be useful when one wants to create and control chiral structures, \textit{e.g.}, homochiral bipolar droplets \cite{Peng2015} or homochiral defect-free double-twist configurations \cite{Eun2019}, because meta-stable configurations may lead to the appearance of unwanted handedness and resultant defects between heterochiral domains. 
Self-assembly is a scalable technique to create chiral meta-materials \cite{Pendry_2004, Soukoulis_2011, Damasceno_2015}, and LCs are popular building blocks for them \cite{Park_2020, Bisoyi_2021}.
Achieving homochirality optimally --- from the choice of materials, dopants, and concentrations to the design of confining geometries and anchoring conditions --- may benefit from the energy-landscape approach. 

\section*{Author contributions}
J. K. conducted the experiments. J. K. and J. J. analyzed the results, performed the numerical calculations, and wrote the manuscript. J. J. supervised the research. All authors discussed the progress of the research and contributed to the final version of the manuscript.

\section*{Conflicts of interest}
There are no conflicts to declare.

\section*{Acknowledgements}
The authors gratefully acknowledge the financial support from the National Research Foundation of Korea through NRF-2021R1A2C101116312 and 2023 Research Fund
(1.230077.01) of Ulsan National Institute of Science and Technology (UNIST).





\bibliography{v2} 
\bibliographystyle{rsc} 

\end{document}


\maketitle

\renewcommand\thefigure{S\arabic{figure}}
\renewcommand{\theequation}{S\arabic{equation}}

\section*{Fingerprint texture at a high concentration of the chiral dopant}

\begin{figure}[h]
\centering
\includegraphics{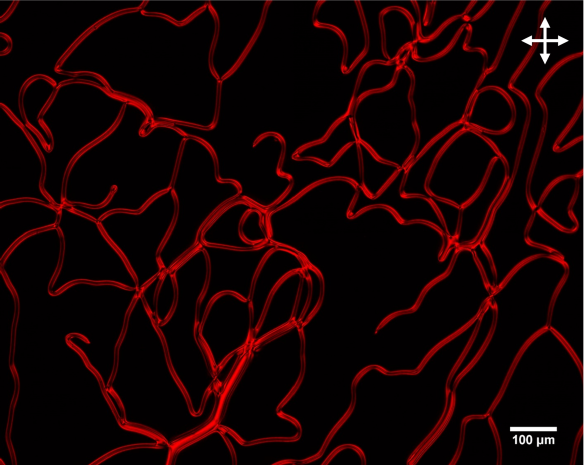}
\caption{
\label{fig:highdenstripe}
The polarized optical microscopy (POM) image of 30.0\% (wt/wt) sunset yellow (SSY) with 0.5\% (wt/wt) brucine sulfate hydrate (BSH) in our sandwich cell at room temperature with crossed polarizers. White double arrows represent the crossed polarizers. The scale bar is 100 $\mu$m.}
\end{figure}

The fingerprint texture appears at high dopant concentrations. Nematic directors at dark regions are homeotropically aligned. In contrast, despite the homeotropic anchoring at the top and bottom substrates, the directors in bright stripes --- the onset of the chiral nematic's fingerprint texture --- are twisted because of the chiral dopants. Thus, we conduct the sessile droplet experiments only at low dopant concentrations where a uniform dark background with no stripes is guaranteed.

\newpage
\section*{Measurement of helical twisting power of brucine sulfate hydrate}

\begin{figure}[h]
\centering
\includegraphics{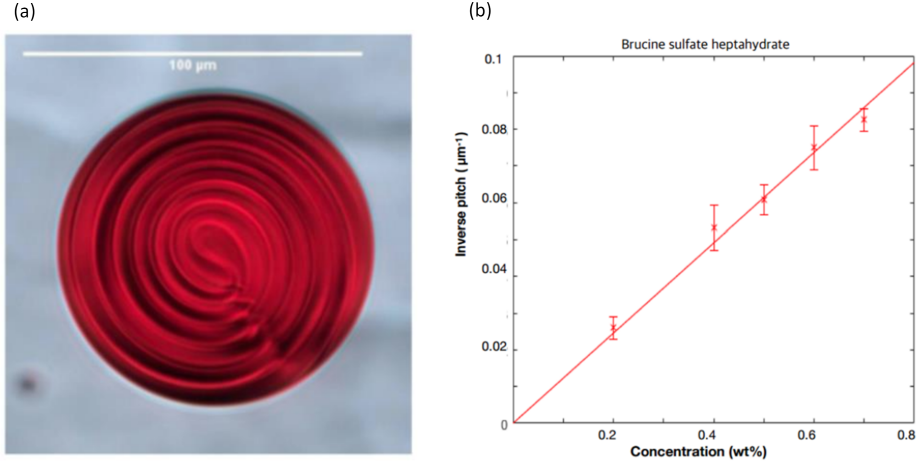}
\caption{
\label{fig:htp}
\textbf{(a)} The bright field microscope image of a droplet of 30.0\% (wt/wt) SSY doped with BSH.
\textbf{(b)} The measured inverse pitches as a function of the chiral dopant concentration.}
\end{figure}

The helical twisting power of BSH in 30.0\% (wt/wt) SSY is measured using fingerprint textures in spherical droplets. As shown in Fig. \ref{fig:htp}(a), a BSH-doped SSY droplet dispersed in silicon oil exhibits the fingerprint texture with a measurable helical pitch. Note that the repeating distance is the half pitch along which the nematic directors rotate by 180 degrees. At various chiral dopant concentrations, the helical pitches are measured, and its inverse pitches are plotted as shown in Fig. \ref{fig:htp}(b). The slope of the linear fit shown as a solid line gives the helical twisting power, 0.123 $\pm$ 0.007 $\mu$m$^{-1}$. This data is adopted from the Master thesis (2019) of Leekyo Jung from Ulsan National Institute of Science and Technology

\newpage
\section*{Rationalization of the director field models}

\begin{figure}[h]
\centering
\includegraphics{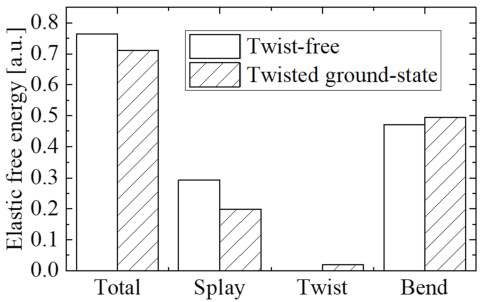}
\caption{
\label{fig:INenergy}
Comparison of elastic free energy between the twist-free planar quadrupole and the energy-minimizing twisted director configuration.
We numerically found the twisted ground state from our ansatz for the I-in-N system, which is introduced in the main text with Eqs. (1-3).
The twisted configuration lowers its total elastic free energy by decreasing splay energy more than the twist and bend penalty.}
\end{figure}

Chiral symmetry breaking in confined LCLC has been reported in the literature. Splay cancellation by the twist deformation is one of the well-known underlying mechanisms for this phenomenon because lyotropic LCs usually have a smaller twist modulus than other moduli. Particularly, when confined, the region near the defects plays a key role because the splay and/or bend deformation is concentrated near them. In other words, twist deformation can also be concentrated near the defect to cancel out the expensive energetic cost of the splay or bend deformation. This inspires us to adopt decaying functions to impose twist deformation to twist-free configuration, \textit{e.g.}, the planar quadrupole configuration. As in Eq. 4 of our main text, we design our twist ansatz so that the degree of the twist deformation decreases as the position gets away from the defect while satisfying the boundary conditions. Specifically, we compare the exponential and power-law decaying functions to find that the power-law function results in a lower total energy (data not shown). Note that this is an improved model compared to the simple linear twist model in Ref. 10. As a result, we show in Fig. \ref{fig:INenergy} that the splay cancellation indeed underlies the chiral symmetry breaking.

\newpage
\section*{Energy landscape in the vicinity of the local minimum in the N-in-I system}
\begin{figure}[h]
\centering
\includegraphics{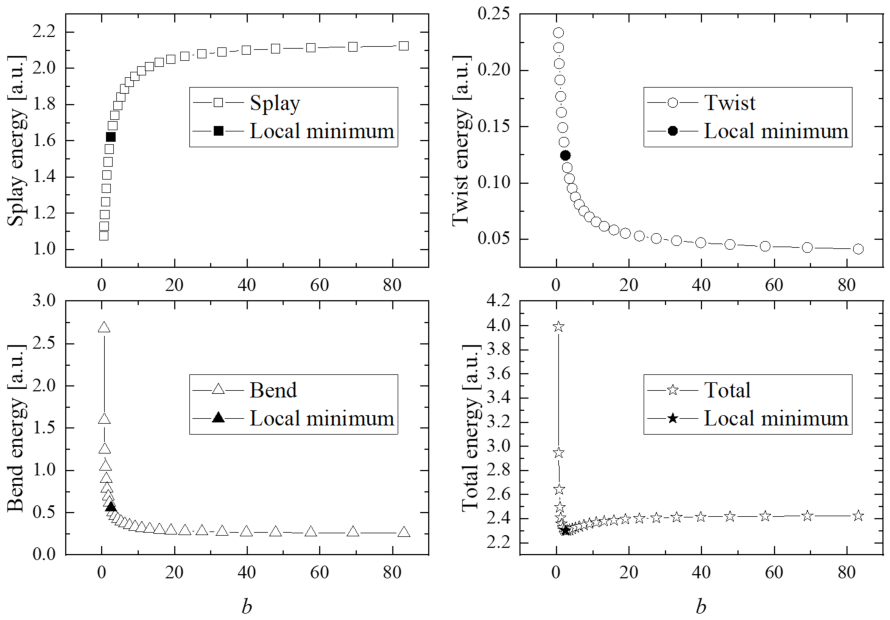}
\caption{
\label{fig:NiIlocalmin}
The energy landscape in the vicinity of the local minimum configuration having the disfavored handedness in the N-in-I system.
To rationalize the existence of the local minimum, we investigate how the energy --- splay, twist, bend, and total, respectively --- changes according to the model parameter $b$ near the local minimum ($a$ = 90, $b$ = 2.51, and $c$ = 0.23) of the system having 0.1\% (wt/wt) of BSH and $K_1 = K_3 = 10~K_2$. This calculation demonstrates that the competition among splay, twist, and bend is responsible for the local minimum, even in the region of the disfavored handedness. Namely, the splay decreases as $b$ gets smaller while twist and bend change in the opposite way, resulting in the local minimum.}
\end{figure}

\newpage
\section*{Determination of critical $k_2$ for the chiral symmetry breaking}

\begin{figure}[h]
\centering
\includegraphics{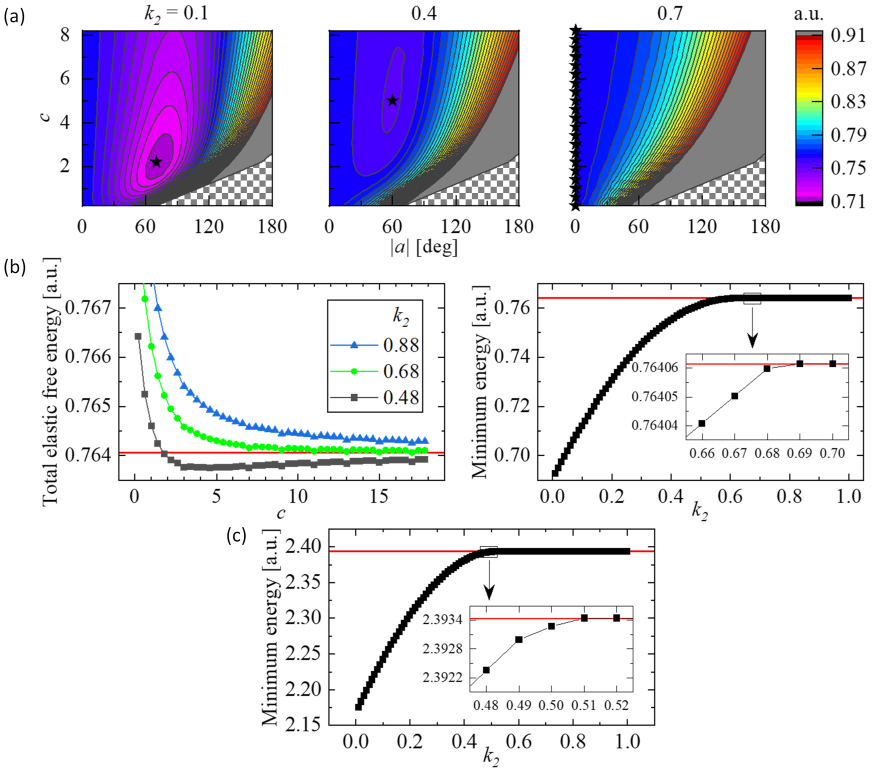}
\caption{
\label{fig:criticalk2}
critical $k_2$ for the chiral symmetry breaking
\textbf{(a)} The elastic free energy landscapes in the I-in-N system according to $k_2 = K_2 / K_3$: 0.1, 0.4 and 0.7, respectively. The filled star symbols indicate the minima.
\textbf{(b)} The determination of the critical $k_2$ for the chiral symmetry breaking in the I-in-N system using two different approaches. The left graph varies the ansatz parameter $c$ at a given $a = 10$ and $k_2$ and compares the total elastic free energy of the twisted configuration with the energy of the twist-free configuration. The right graph plots the smallest total elastic free energy, \textit{i.e.}, the global minimum value in the energy landscape according to $k_2$; the red line is the of the twist-free configuration.
\textbf{(c)} The determination of the critical $k_2$ for the chiral symmetry breaking in the N-in-I system.
}
\end{figure}

In Fig. \ref{fig:criticalk2}(a), we calculate numerically the elastic free energy landscapes in the $a$-$c$ parameter space of the ansatz depending on different $k_{2}=K_{2}/K_{3}$ values: 0.1, 0.4, and 0.7. They show that the twisted ground state with the finite $a$ exists only when $k_{2}$ is small enough; Filled star symbols indicate the energy minima. For instance, with $k_2 = 0.7$, the twist-free director configuration with $a = 0$ is the ground state. This indicates that the chiral-symmetry-breaking transition point exists between $k_2 = 0.4$ and $0.7$. Note that the checkerboard-patterned areas on the right-bottom regions are excluded since director configurations in those regimes considerably violate the boundary conditions.

To narrow down the chiral-symmetry-breaking transition point, we investigate the energy landscape in detail. First, we focus on the shape change from a double-well landscape --- symmetric about $a=0$ with two global minima --- to a single-well shape having one global minimum, which corresponds to the twist-free configuration. To capture this transition while varying $k_2$, we calculate the energy profile along $c$ at the regime next to the symmetric axis. The left graph of Fig. \ref{fig:criticalk2}(b) plots these according to $k_2$. We find that only when $k_2$ becomes smaller than 0.68, there exists at least one $(a,c)$ of which the energy is smaller than the twist-free configuration, which leads to the double-well landscape about $a=0$.

We find the same critical $k_2$ value from another method. We investigate the global minimum energy value over the energy landscape as a function of $k_2$ and find that the minimum value can be smaller than the energy of the twist-free configuration when $k_2 \leq 0.68$, as shown in the right graph of Fig. \ref{fig:criticalk2}(b). As shown in Fig. \ref{fig:criticalk2}(c), we apply the same approach to the N-in-I system to find $k_{\mathrm{crit}} = 0.50$.

\newpage
\section*{Estimation of contact angle and shape of sessile droplets}

\begin{figure}[h]
\centering
\includegraphics{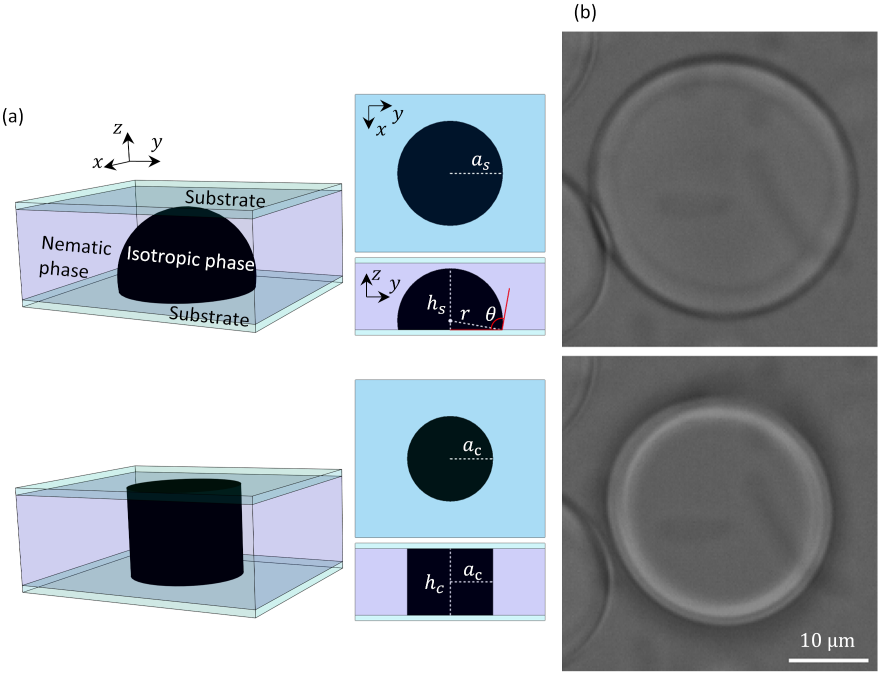}
\caption{
\label{fig:geometryofsystem}
The schematic diagram and the representative experimental results for the estimation of the sessile droplets' contact angle on the parylene-coated glass.
\textbf{(a)} Schematic diagrams of (Top) a sessile droplet sitting at the bottom substrate and  (Bottom)a cylindrical droplet connecting the top and bottom substrates. 
$a_{\mathrm{s}}$ and $h_{\mathrm{s}}$ are the base radius and height of the sessile droplet, respectively. $r$ is the radius of the spherical cap, and $\theta$ is the contact angle. 
$a_{\mathrm{c}}$ and $h_{\mathrm{c}}$ are the base radius and height of the cylindrical droplet, respectively.
\textbf{(b)} Comparison of bright-field images (Top) before and (Bottom) after the transition from the spherical cap to the cylinder. The spherical cap-shaped droplet is surrounded by a dark line, while the cylinder shows a bright boundary.}
\end{figure}

Because the experimental geometry does not allow side-view imaging nor confocal microscopy for the direct imaging of the droplet shape, we estimate indirectly the contact angle utilizing the transition from the spherical-cap droplet to the cylindrical droplet. As shown in Fig. \ref{fig:geometryofsystem}(a), the contact angle $\theta$ is related to the height and radius of the sessile droplet having the spherical-cap shape:
$\theta = 90^{\circ} + \sin^{-1}{\frac{h_{\mathrm{s}} - r}{r}}$, where $h_{\mathrm{s}} > r$ with $\theta \geq$ 90 degrees.
The volume $V_{\mathrm{s}}$ of the spherical cap can be written using $a_{\mathrm{s}}$ and $h_{\mathrm{s}}$: $\frac{1}{6} h_{\mathrm{s}} (3 (\pi a_{\mathrm{s}}^2) + \pi h_{\mathrm{s}}^2) = V_{\mathrm{s}}$.
We measure experimentally the area $\pi a_{\mathrm{s}}^2 = A_{\mathrm{s}}$.
After the transition of the spherical cap to the cylinder spanning the cell gap, we measure experimentally another area $\pi a_{\mathrm{c}}^2 = A_{\mathrm{c}}$.
Assuming the cylinder is straight based on the BF image after the transition, we write the cylinder volume $V_{\mathrm{c}} = (\pi a_{\mathrm{c}}^2) h_{\mathrm{c}}$.
The transition from the spherical cap to the cylinder implies that $h_{\mathrm{s}} \approx h_{\mathrm{c}}$ is approximately the height $h$ of the sandwich cell, and the volume is conserved, \textit{i.e.}, $V_{\mathrm{s}} = V_{\mathrm{c}}$. Thus, we determine the $h = h_{\mathrm{s}} = h_{\mathrm{c}}$ and $r = \sqrt{(h - r)^2 + a_{\mathrm{s}}^2}$, to find $\theta$.
We find the contact angle is slightly greater than $90^{\circ}$: 97.1 and $98.4^{\circ}$ from two independent measurements. The corresponding height $h$ are 18.7 and 21.6 $\mu$m, which are not much different from the thickness of the spacer, 25 $\mu$m.